\documentclass[prd,superscriptaddress,amsfonts,amssymb,amsmath,showpacs,twocolumn]{revtex4-2}
\usepackage{bm}
\usepackage{amsfonts}
\usepackage{latexsym}
\usepackage{graphicx}
\usepackage{amsmath}
\usepackage{palatino}
\usepackage{mathpazo}
\usepackage{textcomp}
\usepackage{float}
\usepackage{booktabs}
\usepackage{dcolumn}
\usepackage{booktabs}
\usepackage{multirow}
\usepackage{hyperref}
\hypersetup{colorlinks,citecolor=blue}
\usepackage{amsmath}
\usepackage{xcolor}
\usepackage{orcidlink}
\usepackage[caption=false]{subfig}
\usepackage{commath}
\def\jnl@style{\it}
\def\aaref@jnl#1{{\jnl@style#1}}
\def\aaref@jnl#1{{\jnl@style#1}}

\allowdisplaybreaks[1]
\renewcommand{\arraystretch}{1.1}
\begin{document}

\color{black}       

\title{Observational constraints on hybrid scale factor in $f(Q,T)$ gravity with anisotropic space-time}

\author{S.A. Narawade\orcidlink{0000-0002-8739-7412}}
\email[Email: ]{shubhamn2616@gmail.com}
\affiliation{Department of Mathematics, Birla Institute of Technology and
Science-Pilani,\\ Hyderabad Campus, Hyderabad-500078, India.}

\author{M. Koussour\orcidlink{0000-0002-4188-0572}}
\email[Email: ]{pr.mouhssine@gmail.com}
\affiliation{Quantum Physics and Magnetism Team, LPMC, Faculty of Science Ben M'sik,\\
Casablanca Hassan II University,
Morocco.} 

\author{B. Mishra\orcidlink{0000-0001-5527-3565}}
\email[Email: ]{bivu@hyderabad.bits-pilani.ac.in}
\affiliation{Department of Mathematics, Birla Institute of Technology and
Science-Pilani,\\ Hyderabad Campus, Hyderabad-500078, India.}


\begin{abstract}
In this paper, we present an accelerating cosmological model by constraining the free parameters using the cosmological datasets in an extended symmetric teleparallel gravity for the flat and anisotropic space-time. We employ a time variable deceleration parameter that behaves early deceleration and late time acceleration in the form of hybrid scale factor (HSF). We obtain the present values of deceleration parameter and analyse the late time behavior of the Universe based on the best-fit values of free parameters. We derive the dynamical parameters of the model and obtain the equation of state parameter at present in the quintessence region; however at late time it approaches to $\Lambda$CDM. The energy conditions are also analysed to validate the modified gravity and we find that strong energy condition is violating. We establish the importance of hybrid scale factor in the late time cosmic phenomena issue.
\end{abstract}

\maketitle

\section{Introduction}\label{sec1}

According to cosmological principle, the perspective of an observer on the Universe is independent of the position and direction of viewing the Universe. So, the present Universe can be modeled through an isotropic and homogeneous space-time such as the FLRW metric. But this may not always be true during the early Universe or in the far future. The classical isotropic and homogeneous model of the Universe needs some fine adjustment as suggested in Wilkinson Microwave Anisotropy Probe (WMAP) \cite{Hinshaw2003, Hinshaw2007, Hinshaw2009}. The inflationary paradigm isotropizes the early Universe to the FLRW geometry as observed now  \cite{Goswami2020}. So, for a complete understanding, spatial inhomogeneity and anisotropy metric should be considered, followed by homogeneity and isotropy. In Bianchi type metrics, one can describe the Universe in the spatially ellipsoidal geometry. For example, Bianchi type-$I$, $V$, and $VII$ are  anistropic and the anisotropy has been eliminated during the inflationary times \cite{Ellis2006, Bonometto2002}, which resulted in an isotropic space-time. The implicit assumption is that the dark energy (DE) is isotropic. So to generate the arbitrary ellipsoidality, the isotropization of the Bianchi metrics can be fine tuned. This is possible if the direction independence on the DE pressure can be relaxed. The Bianchi Universe anisotropy determines Cosmic Microwave Background (CMB) anistropy which needs to be fine tuned. This motivates to study the cosmological model based on the anisotropic space-time, and further the isotropic behaviour can be realized as a special case. \\

The cosmological observations have shown that, the anisotropy in the Universe filled with matter rapidly vanishes and evolves into a FLRW Universe. Bianchi-I Universe also represents an anisotropic Universe. Several authors have studied the Bianchi Type-I Universe in different perspectives.  Using the LRS Bianchi Type-I metric under synchronous gauge and homogeneity bound, it is possible to obtain an anisotropic metric that can be derived from the most general linearly perturbed spatially flat FLRW metric \cite{jcap/AD}. The WMAP data \cite{Hinshaw2003} suggests the similarity to Bianchi morphologies of cosmological models with positive cosmological constant \cite{Jaffe,Jaffe1,Jaffe2,Jaffe3,Jaffe4}. All these studies, someway indicate that the Universe would have developed anisotropic spatial geometry despite inflation, as contrary to the predictions of standard inflationary theories \cite{Guth,Guth1,Guth2,Guth3,Guth4,Guth5,Guth6}. 

The Supernovae Cosmology Project group confirmed the accelerated expansion of the Universe with varieties of explanations \cite{SN1}. The existence of DE has the exotic properties of negative pressure that leads to the negative equation of state (EoS) parameter. The general relativity (GR) has certain issues in addressing this phenomena, hence the modified theories of gravity were explored by altering Einstein-Hilbert action by maintaining the underlying  geometry. Though there are several modifications are available in the literature, the recent non-metricity based gravitational theory provides better insight to the cosmic expansion issue. The symmetric teleparallel theory developed by replacing the Levi-Civita connection of GR with an affine connection. The non-vanishing torsion leads to teleparallel equivalent of GR (TEGR) whereas the non-vanishing non-metricity leads to symmetric telalparalle equivalent of GR (STEGR). Further the $f(Q)$ gravity, where $Q$ be the non-metricity scalar has been proposed \cite{Jimenez2018, Jimenez2020} by extending STEGR. Several works on $f(Q)$ gravity can be found in the literature \cite{Harko2018, Lazkoz2019, Hohmann2019, Soudi2019, Bajardi2020, Lin2021, Frusciante2021, Anagnostopoulos2021, Narawade2022}. Extending further the $f(Q)$ gravity, Xu et al. propose $f(Q,T)$ gravity \cite{Xu2019}, which is a non-minimal coupling between gravity and non-metricity by replacing the Lagrangian with an arbitrary function $f$ and trace of the energy momentum tensor $T$. It is believed that the present cosmic acceleration issue may be clarified by $f(Q,T)$ gravity, which also offers a plausible solution to the DE conundrum \cite {Pati2021, Agrawal2021, Pati2022, Zia2021, Najera2021, Godani2021, Iosifidis2021, Pradhan2021, Najera2022, Shiravand2022, Sokoliuk2022}.

In this paper, we study the anisoropic LRS (Locally Rotationally Symmetric) Bianchi-I space-time with perfect fluid source in $f(Q,T)$ gravity to address the recent cosmic phenomena. Our work represents a significant advance in the study of modified gravity by exploring the behavior of the $f(Q,T)$ theory at the background of an anisotropic space-time. Though there is no restriction on the kind of space-time used for the cosmology study, but commonly, for the late time phenomena the FLRW space-time are considered whereas for black hole, worm hole study the static spherically symmetric are most appropriate. At the same time the Bianchi type anisotropic space-time are considered for the study of the problems related to the early universe and so on. However relatively not much exploration are done in recent research on the late time cosmic phenomena in an anistropic space-time. In this study, we have given a detailed formulation of the $f(Q,T)$ theory in an anisotropic space-time. The aim of the study is to understand and analyse the accelerating behaviour of the cosmological model in an anisotropic background with the use of cosmological date sets, as the expansion in spatial directions are not uniform.\\

We discuss some recent research based on the anisotropic space-time in recent years. Esposito et al. \cite{espo} have presented a reconstruction algorithm for cosmological models based on $f(Q)$ gravity; specifically the focus was on to obtain the exact solutions of the field equations and its application in the spontaneous isotropization of Bianchi type-I models. The behavior of viscous DE in Bianchi type-V spacetime has been studied by Hassan \cite{A2}. Also Hassan \cite{A4} has constrained the parameters of the Bianchi type-I DE model, where the expansion rate of the Universe as $67.9\pm 1.2$ was obtained. The Markov chain Monte Carlo method to place constraints on the cosmological parameters of five DE models including flat and curved $X$CDM, flat and curved $\Lambda$CDM and Bianchi type-I spacetime has been analysed by Hassan and Soroush \cite{A5}. Most of the research in an anisotropic space-time explored the model for the late time accelerating phase only. In this work, we will study the models with a time varying deceleration parameter that shows the decelerating behaviour at the early Universe and accelerating behaviour at late times.\\

The paper is organised as follows: In Sec. \ref{sec2}, a complete discussion of $f(Q,T)$ gravity and the gravitational action with general field equations are presented. In Sec. \ref{sec3}, the field equations of of $f(Q,T)$ gravity in LRS Bianchi-I space-time has been derived and the dynamical parameters are expressed in Hubble term. The observational analysis has been performed for hybrid scale factor using $Hubble$, $Pantheon$ and $BAO$ data set in Sec. \ref{sec4}. The DE cosmological parameters are analysed in Sec. \ref{sec5}. Finally, in Sec. \ref{sec6}, we have concluded with the results and discussions.

\section{$f(Q,T)$ gravity field equations} \label{sec2} 
The $f(Q,T)$ gravity is constrained with the curvature and torsion free assumptions, i.e. $R^{\rho }{}_{\sigma \mu \nu }=0$ and $T^{\rho }{}_{\mu \nu }=0$. The corresponding connection, known as the Weyl connection, is defined as $\widetilde{\Gamma }{^{\sigma }}_{\mu \gamma }=L^{\lambda }{}_{\mu \nu}+\Gamma {^{\sigma }}_{\mu \gamma }$, where $L^{\lambda }{}_{\mu\nu }$ and $\Gamma {^{\sigma }}_{\mu \gamma }$ are the disformation tensor and the Levi-Civita connection, respectively. For the $f(Q,T)$ gravity, the
Modified Einstein-Hilbert (MEH) action is given by \cite{Xu2019}, 
\begin{equation}  \label{1}
S_{MEH}=\int \sqrt{-g}\left[ \frac{1}{16\pi }f(Q,T)+L_{m}\right] d^{4}x,
\end{equation}
where $f(Q,T)$ is an arbitrary function of the non-metricity scalar $Q$ and $T$ is the trace of the energy-momentum tensor ($T_{\mu \nu }$); and $g$ is the determinant of the metric tensor $g_{\mu \nu }$. The energy-momentum tensor $T_{\mu \nu }$ from the Lagrangian
matter is,
\begin{equation}
T_{\mu \nu }=-\frac{2}{\sqrt{-g}}\dfrac{\delta (\sqrt{-g}L_{m})}{\delta
g^{\mu \nu }}.
\end{equation}
The variation of the energy-momentum tensor with respect to the metric
tensor yields, 
\begin{equation}  \label{10}
\frac{\delta \,g^{\,\mu \nu }\,T_{\,\mu \nu }}{\delta \,g^{\,\alpha \,\beta }%
}=T_{\,\alpha \beta }+\Theta _{\,\alpha \,\beta }\,,
\end{equation}
and 
\begin{equation}
\Theta _{\mu \nu }=g^{\alpha \beta }\frac{\delta T_{\alpha \beta }}{\delta
g^{\mu \nu }}.  \label{9}
\end{equation}
The expression for the non-metricity scalar $Q$ is \cite{Jimenez2018,Jimenez2020}, 
\begin{equation}
Q\equiv -g^{\mu \nu }(L_{\,\,\,\alpha \mu }^{\beta }L_{\,\,\,\nu \beta
}^{\alpha }-L_{\,\,\,\alpha \beta }^{\beta }L_{\,\,\,\mu \nu }^{\alpha }),
\label{2}
\end{equation}%
where 
\begin{equation}
L_{\alpha \gamma }^{\beta }=\frac{1}{2}g^{\beta \eta }\left( Q_{\gamma
\alpha \eta }+Q_{\alpha \eta \gamma }-Q_{\eta \alpha \gamma }\right) ={%
L^{\beta }}_{\gamma \alpha }.  \label{3}
\end{equation}

and the non-metricity tensor, 
\begin{equation}
Q_{\gamma \mu \nu }=-\nabla _{\gamma }g_{\mu \nu }=-\partial _{\gamma
}g_{\mu \nu }+g_{\nu \sigma }\widetilde{\Gamma }{^{\sigma }}_{\mu \gamma
}+g_{\sigma \mu }\widetilde{\Gamma }{^{\sigma }}_{\nu \gamma },  \label{4}
\end{equation}%
The trace of the non-metricity tensor becomes, 
\begin{equation}
Q_{\beta }=g^{\mu \nu }Q_{\beta \mu \nu },\qquad \widetilde{Q}_{\beta
}=g^{\mu \nu }Q_{\mu \beta \nu }.  \label{5}
\end{equation}%
The superpotential tensor, known as the non-metricity conjugate is
defined as, 
\begin{eqnarray}
\hspace{-0.5cm} &&P_{\ \ \mu \nu }^{\beta }\equiv \frac{1}{4}\bigg[-Q_{\ \
\mu \nu }^{\beta }+2Q_{\left( \mu \ \ \ \nu \right) }^{\ \ \ \beta
}+Q^{\beta }g_{\mu \nu }-\widetilde{Q}^{\beta }g_{\mu \nu }  \notag \\
\hspace{-0.5cm} &&-\delta _{\ \ (\mu }^{\beta }Q_{\nu )}\bigg]=-\frac{1}{2}%
L_{\ \ \mu \nu }^{\beta }+\frac{1}{4}\left( Q^{\beta }-\widetilde{Q}^{\beta
}\right) g_{\mu \nu }-\frac{1}{4}\delta _{\ \ (\mu }^{\beta }Q_{\nu )}.\quad
\quad   \label{6}
\end{eqnarray}

Hence, the scalar of non-metricity yields \cite{Jimenez2018}, 
\begin{eqnarray}
&&Q=-Q_{\beta \mu \nu }P^{\beta \mu \nu }=-\frac{1}{4}\big(-Q^{\beta \nu
\rho }Q_{\beta \nu \rho }+2Q^{\beta \nu \rho }Q_{\rho \beta \nu }  \notag \\
&&-2Q^{\rho }\tilde{Q}_{\rho }+Q^{\rho }Q_{\rho }\big).  \label{7}
\end{eqnarray}

The $f(Q,T)$ gravity field equations are derived by varying the action \eqref{1} with respect to the metric component as, 
\begin{multline}
\frac{2}{\sqrt{-g}}\nabla _{\beta }(f_{Q}\sqrt{-g}P_{\,\,\,\,\mu \nu
}^{\beta })-\frac{1}{2}fg_{\mu \nu }+f_{T}(T_{\mu \nu }+\Theta _{\mu \nu })
\label{11} \\
+f_{Q}(P_{\mu \beta \alpha }Q_{\nu }^{\,\,\,\beta \alpha }-2Q_{\,\,\,\mu
}^{\beta \alpha }P_{\beta \alpha \nu })=8\pi T_{\mu \nu }.
\end{multline}
where $f_{Q}\equiv f_{Q}\left( Q,T\right)$ and $f_{T}\equiv f_{T}\left(Q,T\right)$ are the partial derivatives of $f\equiv f\left(Q,T\right)$ with respect to $Q$ and $T$ respectively. It is important to mention here that the above field equation differs from that of \cite{Xu2019} due to the sign used in defining the scalar of non-metricity. However, it does not have any effect on the correctness and results. We consider the matter content of the Universe is that of perfect fluid. So, the energy-momentum tensor for the perfect fluid is, $T_{\mu \nu }=\left( \rho +p\right) u_{\mu }u_{\nu }+pg_{\mu \nu}$, where $\rho$, $p$, and $u^{\mu }$ respectively represents the energy density, pressure, and 4-vector velocity of the perfect fluid. The matter Lagrangian,  $L_{m}=p$ that depends only on the metric tensors and hence $\Theta _{\mu \nu }=pg_{\mu \nu }-2T_{\mu \nu }$.

\section{Field Equations of LRS Bianchi-I cosmology in $f(Q,T)$ gravity}\label{sec3} 
We consider an anisotropic metric in the form of LRS Bianchi-I space-time as, 
\begin{equation}
ds^{2}=-dt^{2}+A^{2}(t)dx^{2}+B^{2}(t)(dy^{2}+dz^{2}),
\end{equation}
where, $A\left( t\right)$ represents the expansion in $x$-direction whereas $B\left( t\right)$ in $y$, and $z$ directions. If $A\left( t\right) =B\left( t\right)=a\left( t\right) $, then it reduces to the flat FLRW cosmology. To obtain the motion equations of a test particle in this
coincident gauge choice, we employ the standard flat affine connection. So, the the non-metricity scalar for LRS Bianchi-I cosmology is,
\begin{equation}
Q=-6(2H-H_{y})H_{y}\,,  \label{Q}
\end{equation}
where%
\begin{equation}
H_{x}=\frac{\dot{A}}{A},\quad H_{y}=\frac{\dot{B}}{B},\quad H_{z}=H_{y}\,,
\end{equation}
and%
\begin{equation}
H=\frac{1}{3}\frac{\dot{V}}{V}=\frac{1}{3}\left[ H_{x}+2H_{y}\right] \,,
\label{H}
\end{equation}
represent the directional Hubble parameters along the associated coordinate axes and the average Hubble parameter, respectively. If $H_{x}=H_{y}=H$ , the equation $Q=-6(2H-H_{y})H_{y}$ reduces to $Q=-6H^{2}$, represents the isotropic case. The spatial volume becomes, 
\begin{equation}
V=AB^{2}=a(t)^{3}.  \label{V}
\end{equation}

The average anisotropic parameter $\Delta $ of the expansion used to determine whether the model approaches isotropy or not, can be expressed as,  
\begin{equation}
\Delta =\frac{1}{3}\sum_{i=1}^{3}\left( \frac{H_{i}-H}{H}\right) ^{2}=\frac{2%
}{9H^{2}}\left( H_{x}-H_{y}\right) ^{2}.  \label{delta}
\end{equation}%
where $H_{i}\left( i=x,y,z\right) $. It is important to note that the
deviation from isotropic expansion is measured by $\Delta $, and the Universe expands isotropically when $\Delta =0$. In addition, the expansion scalar $\theta $ and shear $\sigma $ can be expressed in the form of Hubble parameter as, 
\begin{equation}
\theta =H_{x}+2H_{y},\quad \sigma =\frac{|H_{x}-H_{y}|}{\sqrt{3}}.
\label{shear}
\end{equation}

The deceleration parameter is given by%
\begin{equation}\label{eq:19}
q=-1+\frac{d}{dt}\left( \frac{1}{H}\right) .
\end{equation}

The field equations Eq. \eqref{11} reduces to \cite{Loo2023},
\begin{widetext}
\begin{align}
(8\pi +f_{T})\rho +f_{T}p=& \frac{f}{2}+6f_{Q}(2H-H_{y})H_{y}\,,  \label{rho}
\\
8\pi p=& -\frac{f}{2}-\frac{\partial }{\partial t}\left[ 2f_{Q}H_{y}\right]
-6f_{Q}H_{y}H,  \label{p_x} \\
8\pi p=& -\frac{f}{2}-\frac{\partial }{\partial t}\left[ f_{Q}(3H-H_{y})%
\right] -3f_{Q}(3H-H_{y})H\,.  \label{p_y}
\end{align}

An algebraic manipulations between Eqs. (\ref{rho})-(\ref{p_y}), the field equations can be reduced to, 
\begin{align}
8\pi \rho =& \frac{f}{2}+\frac{6f_{Q}}{8\pi +f_{T}}\left[ 8\pi
(2H-H_{y})H_{y}+f_{T}H^{2}\right] +\frac{2f_{T}}{8\pi +f_{T}}\frac{\partial 
}{\partial t}\left[ f_{Q}H\right] \,,  \label{rho-2} \\
8\pi p=& -\frac{f}{2}-2\frac{\partial }{\partial t}\left[ f_{Q}H\right]
-6f_{Q}H^{2}\,.  \label{p-2}
\end{align}
\end{widetext}

Theoretically, we are interested in integrating the field equations (\ref{rho-2})-(\ref{p-2}) while taking Eqs. (\ref{Q}) and (\ref{H}) into consideration. However, because of the complexity involved, it would be difficult to solve Eqs. (\ref{rho-2})-(\ref{p-2}), hence we shall introduce additional constraints into the system of equations. These constraints and proposed solutions will be discussed in the subsequent section.

\subsection{The $f\left( Q,T\right) =Q+bT$\ Model}

The current research will concentrate on the cosmological features of $f(Q,T) $ theory, with the simple relationship between $Q$ and $T$ as, $f\left( Q,T\right) =Q+bT$, where $b$ is the model parameter. Then,  $f_{Q}=1$ and $f_{T}=b$. This specific form of $f\left(Q,T\right) $ is similar to the linear functional form discussed in \cite{Xu2019}. To note, $b=0$ leads to the equivalent scenario of GR, which has been well justified in Ref. \cite{Loo2023}. The functional of $f(Q,T)$ was thoroughly examined in Ref. \cite{Xu2019} which demonstrates that the evolution of Universe has undergone accelerating expansion and will be terminating with the de-Sitter type evolution. Now, Eq. (\ref{rho-2}) and Eq. (\ref{p-2}) becomes, 
\begin{eqnarray}
\rho &=& \frac{3H^{2}+\overset{.}{H}}{2b+8\pi }-\frac{3(H-H_{y})^{2}+\overset%
{.}{H}}{b+8\pi }~, \\
p &=& -\frac{3H^{2}+\overset{.}{H}}{2b+8\pi }-\frac{3(H-H_{y})^{2}+\overset{.%
}{H}}{b+8\pi }.
\end{eqnarray}
and the equation of state (EoS) parameter $\omega =\frac{p}{\rho }$, becomes, 
\begin{widetext}
\begin{equation}
\omega =\frac{3b\left( 3H^{2}-4HH_{y}+2H_{y}^{2}+\overset{.}{H}\right) +8\pi
\left( 6H^{2}-6HH_{y}+3H_{y}^{2}+2\overset{.}{H}\right) }{b\left( 3\left(
H^{2}-4HH_{y}+2H_{y}^{2}\right) +\overset{.}{H}\right) +24\pi H_{y}(H_{y}-2H)
}.
\end{equation}
\end{widetext}

To investigate the behavior of the aforementioned parameters, we must first evaluate the constraining relation. To begin, we will assume an anisotropic relation that can be expressed in terms of shear ($\sigma $) and expansion scalar ($\theta $) as $\sigma ^{2}\propto \theta ^{2}$. This implies that while $\frac{\sigma }{\theta }$ is constant, the Hubble expansion of the Universe can attain isotropic \cite{Ani0, Ani1, Ani2, Ani3, Koussour1, Koussour2, Koussour3}. This condition leads to $A\left( t\right) =B\left(t\right) ^{\gamma }$, where $\gamma \neq 1$ is an arbitrary constant. When $\gamma =1$, the model becomes isotropic. This yields the relationship between directional Hubble parameters as,
\begin{equation}
H_{x}=\gamma H_{y}.
\end{equation}
The average Hubble parameter becomes,
\begin{equation}
H=\frac{\left( \gamma +2\right) }{3}H_{y}.
\end{equation}

Now, we assume the average scale factor to be a combination of power-law (PL) and exponential law (EL), $a\left( t\right) =t^{\alpha }e^{\chi t}$, where $\alpha $ and $\chi $ are non-negative constants \cite{Mishra,Akarsu}. This relationship yields the EL when $\alpha =0$ and the PL when $\chi =0$. This is referred to as the Hybrid Scale Factor (HSF), which combines PL and EL expansions. Using this average scale factor, we get a time-dependent deceleration parameter. Some more analysis on HSF in GR and MGT can be seen in Refs. \cite{HEL0,HEL1,Mishra2018,HEL2, HEL3,HEL4}. At present time ($t=t_{0}$), the average
scale factor is expressed in the form $a\left( t_{0}\right)=a_{0}=t_{0}^{\alpha }e^{\chi t_{0}}$, where $a_{0}$ is the present value of the scale factor. Hence, the HSF can be written in terms of the present scale factor and the age of the Universe as $a\left( t\right) =a_{0}\left(\frac{t}{t_{0}}\right)^{\alpha}e^{\beta \left(\frac{t}{t_{0}}-1\right)}$, where $\beta =\chi t_{0}\geq 0$. The average Hubble parameter and the deceleration parameter are cosmological parameters that are determined as $H=\frac{\alpha}{t}+\frac{\beta }{t_{0}}$, and $q=-1+\frac{\alpha t_{0}^{2}}{\left(\alpha t_{0}+\beta t\right)^{2}}$.

It is obvious that the constants $\alpha$ and $\beta$ can be chosen so that the PL term dominates over the EL term in the early Universe, i.e. at the time scales of primordial nucleosynthesis. As a result, at $t\sim 0$, the scale factor parameters become, $H\sim \frac{\alpha }{t}$, and $q\sim -1+\frac{1}{\alpha }$. Again, at late time, the EL term dominates, such that in the limit $t\rightarrow \infty$, we obtain $H\rightarrow \frac{\beta }{t_{0}}$, and $q\rightarrow -1$. It can be shown that the parameter $\alpha$ governs early phase of the Universe, whereas the parameter $\beta $ governs the late phase of the Universe \cite{Mishra}. Since $\alpha$ and $\beta$ are both non-zero, the Universe evolves with the variable deceleration parameter, and the transition from deceleration to acceleration occurs at $t=\frac{t_{0}}{\beta }\left( \sqrt{\alpha }-\alpha \right)$, keeping $\alpha$ in the range $0<\alpha<1$.

Using the relationship between the scale factor and redshift $\frac{a_{0}}{a}=1+z$, where we have assumed that the scale factor for the current epoch is 1, we can obtain the relationship between $t$ and $z$ as $t\left(z\right) =\left( \frac{\alpha t_{0}}{\beta }\right) \psi \left(
z\right) $, where $\psi \left( z\right) =W\left[ \frac{\beta }{\alpha }e^{\frac{\beta-\ln \left( 1+z\right) }{\alpha }}\right]$, $W$ represents the Lambert function. So, the Hubble parameter  $H\left(z\right) $ in terms of redshift $z$ becomes, 
\begin{equation}
H\left( z\right) =\frac{H_{0}\beta }{\alpha +\beta }\left[ \frac{\psi \left(
z\right) +1}{\psi \left( z\right) }\right] ,  \label{Hz}
\end{equation}
where $H_{0}=\frac{\alpha +\beta }{t_{0}}$ is the present Hubble value ($z=0$). Now, the goal is to determine the values of the model parameters ($H_{0}$, $\alpha $, and $\beta $), the dynamical parameters ($\rho $, $p$, $\omega $) using different types of observational data. This approach allows us to construct a cosmological model that is consistent with the available astrophysical observations and is physically plausible. By comparing the model predictions with observations, we can constrain the values of the parameters and ensure that the model accurately describes the behavior of the Universe. This will help us in the better understanding of the properties and evolution of the Universe. 

\section{Observational constraints}
\label{sec4} 
We employ the recent cosmological datasets to constrain the model parameters of $H_{0}$, $\alpha $ and $\beta $ as given below:
\begin{itemize}
\item \textbf{Hubble $H(z)$}: We analyze $Hubble$ data points obtained using the differential age approach \cite%
{Yu/2018,Moresco/2015,Sharov/2018}.

\item \textbf{Baryon Acoustic Oscillation (BAO)}: We also analyze the BAO data from the SDSS-MGS, Wiggle Z, and 6dFGS experiments \cite%
{Blake/2011,Percival/2010,Giostri/2012}.

\item \textbf{Type-Ia Supernova measurement (SNe Ia)}: Finally, we look at 1048 SNe Ia luminosity distance estimates from the Pan-STARSS 1 (PS1) Medium Deep Survey, the Low-z, SDSS, SNLS, and HST missions in the Pantheon sample \cite{Scolnic/2018,Chang/2019}.
\end{itemize}

We introduce the novel approach to analyze the cosmological observational data using the Markov Chain Monte Carlo (MCMC) sampling technique. Though this method has been used extensively in the literature, we present here a comprehensive analysis that includes extensive set of data and more stringent set of priors on the model parameters. To be specific, we focus on the parameter space $\theta_{s}=(H_{0},~\alpha,~\beta)$ and use the publicly available Python package \textit{emcee} \cite{Mackey/2013}, to parallelize the MCMC sampling process. Our analysis employs three types of data: $Hubble$, $BAO$, and $SNe$ Ia data. We test the priors for the parameters as $60.0<H_{0}<80.0$, $0.0<\alpha<10.0$, and $0.0<\beta<10.0$. We use 100 walkers and 1000 steps to determine the results of our MCMC study.

\subsection{$Hubble$ data}

We employ a collection of $Hubble$ data points composed of 37 points collected using the differential age (DA) approach to estimate the expansion rate of the Universe at redshift $z$. Hence, the Hubble parameter $H(z)$ can be written as 
\begin{equation}\label{eq:31}
H(z)= -\frac{1}{1+z} \frac{dz}{dt}.
\end{equation}

To get the mean values of the model parameters $\theta_{s}=(H_{0}, \alpha, \beta)$, we applied the chi-square function ($\chi^{2}$) for $Hubble$ data as, 
\begin{equation}
\chi^{2}_{H} = \sum_{i=1}^{37} \frac{\left[H(\theta_{s}, z_{i})-
H_{obs}(z_{i})\right]^2}{\sigma(z_{i})^2}.
\end{equation}

Here, $H(z_{i})$ indicates the theoretical value for a specific model (in this case, HSF) at various red-shifts $z_{i}$, and $H_{obs}(z_{i})$ indicates the observational
value, $\sigma(z_{i})$ indicates the observational error.

\subsection{$BAO$ data}

For $BAO$ data, we use a combination of SDSS, 6dFGS, and Wiggle Z surveys at various red-shifts. This research combines $BAO$ data as well as the following cosmology: 
\begin{equation}
d_{A}(z)=c \int_{0}^{z} \frac{dy}{H(y)},
\end{equation}
\begin{equation}
D_{v}(z)=\left[ \frac{d_{A}^2 (z) c z }{H(z)} \right]^{1/3}.
\end{equation}

Here, $d_{A}(z)$ indicates the comoving angular diameter distance, and $%
D_{v}$ indicates the dilation scale. Furthermore, for $BAO$ data, the chi-square function ($
\chi^{2}$) becomes,
\begin{equation}
\chi^{2}_{BAO} = X^{T} C_{BAO}^{-1} X,
\end{equation}
where $X$ is determined by the survey under consideration and $C_{BAO}^{-1}$ indicates the
inverse covariance matrix \cite{Giostri/2012}.

\subsection{$SNe$ data}

To get the best results with $SNe$ Ia data, we start with the observed distance modulus $\mu_{obs}$ generated by $SNe$ Ia detections and compare it to the theoretical value $\mu_{th}$. The Pantheon sample, a recent $SNe$ Ia dataset including 1048 distance modulus $\mu_{obs}$ at different red-shifts in the range $0.01<z<2.26$, is used in this work. The distance modulus from each $SNe$ can be determined using the following equations:
\begin{equation}
\mu_{th}(z)=5 log_{10} \frac{d_{l}(z)}{Mpc}+25,
\end{equation}
\begin{equation}
d_{l}(z)=c(1+z) \int_{0}^{z} \frac{dy}{H(y,\theta)},
\end{equation}
where $c$ indicates the velocity of light. The following relationship can be applied to compute the distance modulus, 
\begin{equation}
\mu= m_{B}-M_{B}+\alpha x_{1} - \beta\, c + \Delta_{M} + \Delta_{B}.
\end{equation}

Here, $m_{B}$ indicates the measured peak magnitude at the $B$-band maximum and $%
M_{B}$ indicates the absolute magnitude. The parameters $c$, $\alpha$, $\beta$, and 
$x_{1}$, respectively, conform to the color at the brightness point, the
luminosity stretch-color relation, and the light color shape. Furthermore, $%
\Delta_{M}$ and $\Delta_{B}$ indicate distance adjustments based on the host
galaxy's mass and simulation-based anticipated biases. The nuisance parameters can be calculated using a unique approach called as BEAMS with Bias Corrections (BBC) \cite{Kessler/2017}. Hence, the measured distance modulus equals the difference between the apparent $m_{B}$ and absolute $M_{B}$ magnitudes, i.e. $\mu = m_{B}-M_{B}
$. For the $SNe$ data, the $\chi^{2}$ function is,

\begin{equation}
\chi^{2}_{SNe} =\sum_{i,j=1} ^{1048} \Delta \mu_{i} \left(
C_{SNe}^{-1}\right)_{ij} \Delta \mu_{j},
\end{equation}
where $\Delta \mu_{i}= \mu_{th}-\mu_{obs}$ and $C_{SNe}$ indicates the
covariance matrix.  The profile of our model versus $SNe$ data is displayed in Fig.- \ref{ErrorSNe}. The figure also shows a comparison of our model to the commonly
used $\Lambda$CDM model in cosmology. Also, our model closely matches the observed SNe data.

Now, the $\chi^{2}$ function for the $Hubble+BAO+Pantheon$ data is considered to be, 
\begin{equation}
\chi^{2}_{total}=\chi^{2}_{H}+\chi^{2}_{BAO}+\chi^{2}_{SNe},
\end{equation}

By using the above $Hubble$ data, we obtained the best-fit values of the model parameters $H_{0}$, $\alpha$, and $\beta$, as shown in Fig.- \ref{F_H} with $1-\sigma$ and $2-\sigma$ likelihood contours. The best-fit values obtained are $H_{0}=67.7^{1.7}_{-1.7}$, $\alpha=0.57^{+0.14}_{-0.13}$, and $\beta=0.38^{+0.12}_{-0.13}$. The profile of our model versus $Hubble$ data is displayed in Fig.- \ref{ErrorHubble}. The figure also shows a comparison of our model to the commonly used $\Lambda$CDM model in cosmology i.e. $H\left( z\right) =H_{0}\sqrt{\Omega _{m}^{0}\left( 1+z\right) ^{3}+\Omega _{\Lambda}^{0}}$ (we have considered $\Omega _{m}^{0}=0.3$, $\Omega _{\Lambda}^{0}=0.7$) \cite{Planck2020}. As observed, the model closely matches the observed $Hubble$ data. Using the combined $Hubble+BAO+Pantheon$ data, we obtained the best-fit values of the model parameters $H_{0}$, $\alpha$, and $\beta$, as shown in Fig.- \ref{F_total} with $1-\sigma$ and $2-\sigma$ likelihood contours. The best-fit values obtained are $H_{0}=67.2^{+1.2}_{-1.2}$, $
\alpha=0.603^{+0.028}_{-0.030}$, and $\beta=0.355^{+0.044}_{-0.043}$. Also, the current values for various cosmological parameters $H_{0}$, $q_{0}$, $z_{tr}$ and $\omega _{0}$ are summarized in Table- \ref{tab}.\\

\begin{figure}[h]
\centerline{\includegraphics[scale=0.72]{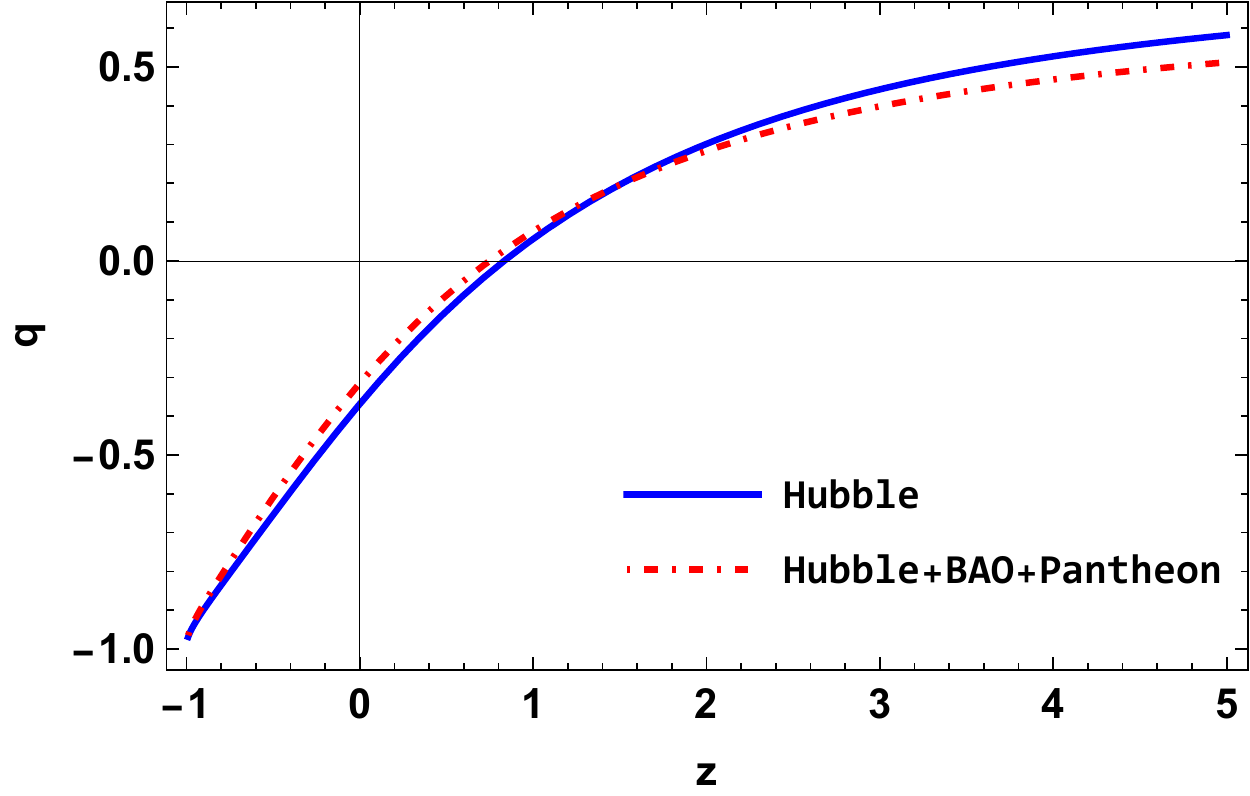}}
\caption{Plot of $q$ versus $z$.}
\label{F_q}
\end{figure}

From Eq. \eqref{eq:19} and Eq.\eqref{eq:31}, we have $q(z) = -1+\frac{(1+z)H_{z}(z)}{H(z)}$. A given interval on $q$ ($q_{0}$ is the present value) describes how the Universe behaves: the Universe experiences expanding behavior and undergoes deceleration phase when $q_{0}>0$, while the Universe is expanding and accelerating at the present time for $-1 < q_{0} < 0$. As a result of inflation during the very early Universe known as the de-Sitter Universe, $q_{0} = -1$ represents the Universe at present. We obtained the present value of $q(z)$ from $Hubble$ and $Hubble+BAO+Pantheon$ datasets respectively $q_{0}=-0.37^{+0.15}_{-0.15}$, and $q_{0}=-0.343^{+0.062}_{-0.061}$ [FIG.- \ref{F_q}], which shows that the Universe undergoes accelerated expansion \cite{Bonilla2018, Giostri/2012}. Also we obtained the value of transition from deceleration to acceleration for hybrid scale factor using $Hubble$ and $Hubble+BAO+Pantheon$ dataset as $z_{t}=0.83^{+1.87}_{-1.87}$ and $z_{t}=0.75^{+0.36}_{-0.37}$ respectively  which suited with the observational values \cite{Capozziello2014,  Capozziello2015,  Farooq2017}.

\begin{widetext}

\begin{figure}[h]
\centerline{\includegraphics[scale=0.7]{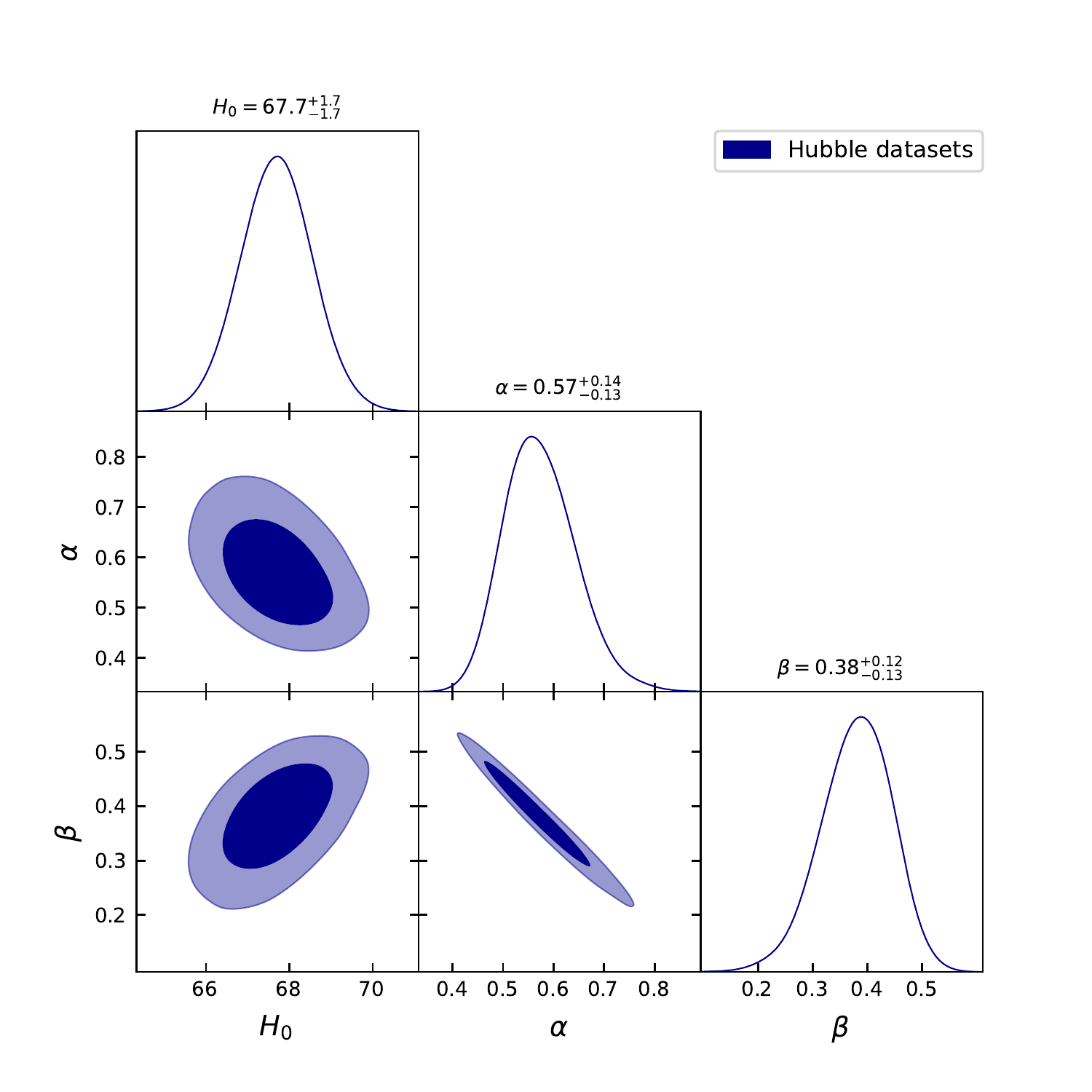}}
\caption{The $1-\sigma$ and $2-\sigma$ confidence curves for the model parameters $H_{0}$, $\alpha$, and $\beta$ with $Hubble$ data.}
\label{F_H}
\end{figure}

\begin{figure}[h]
\centerline{\includegraphics[scale=0.60]{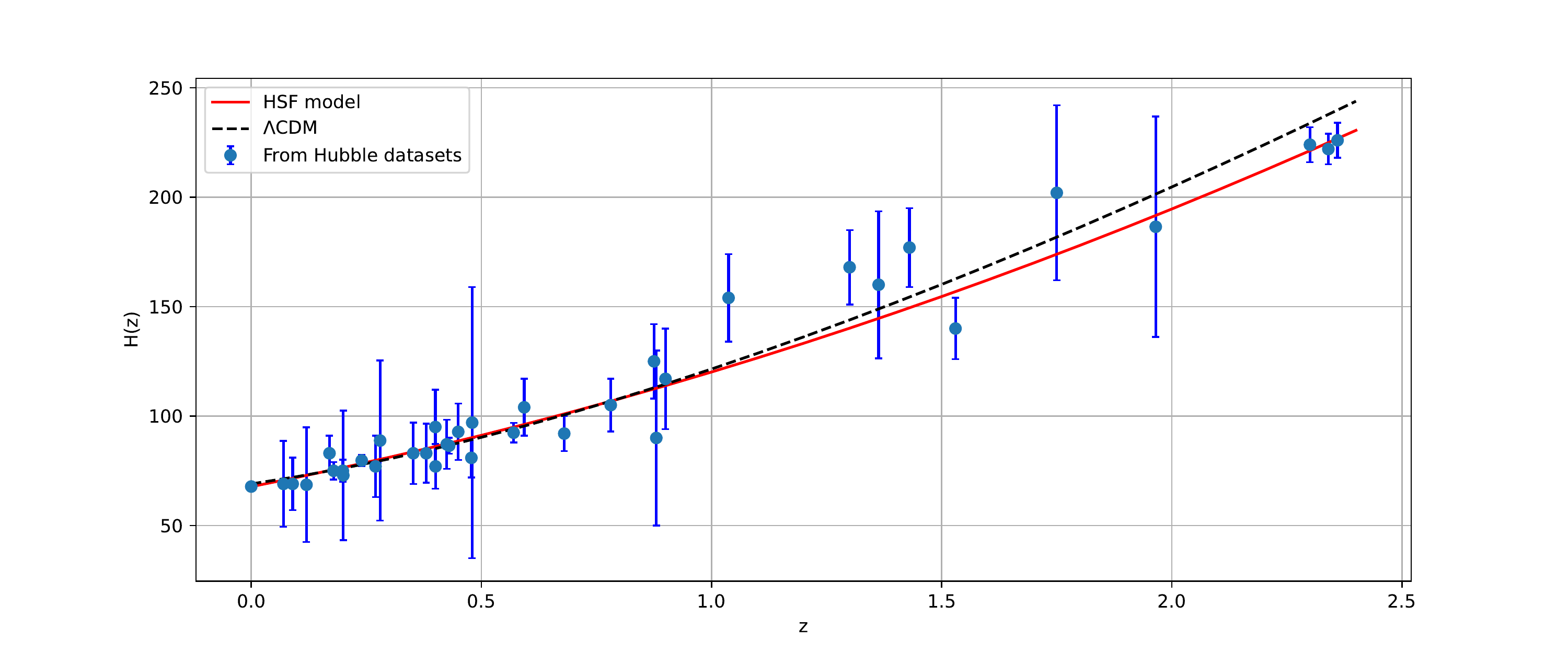}}
\caption{Evolution of $H(z)$ in red-shift $z$. The black dashed line represents the $\Lambda$CDM model, the red line is the HSF model curve, and the blue dots show error bars.}
\label{ErrorHubble}
\end{figure}

\begin{figure}[h]
\centerline{\includegraphics[scale=0.60]{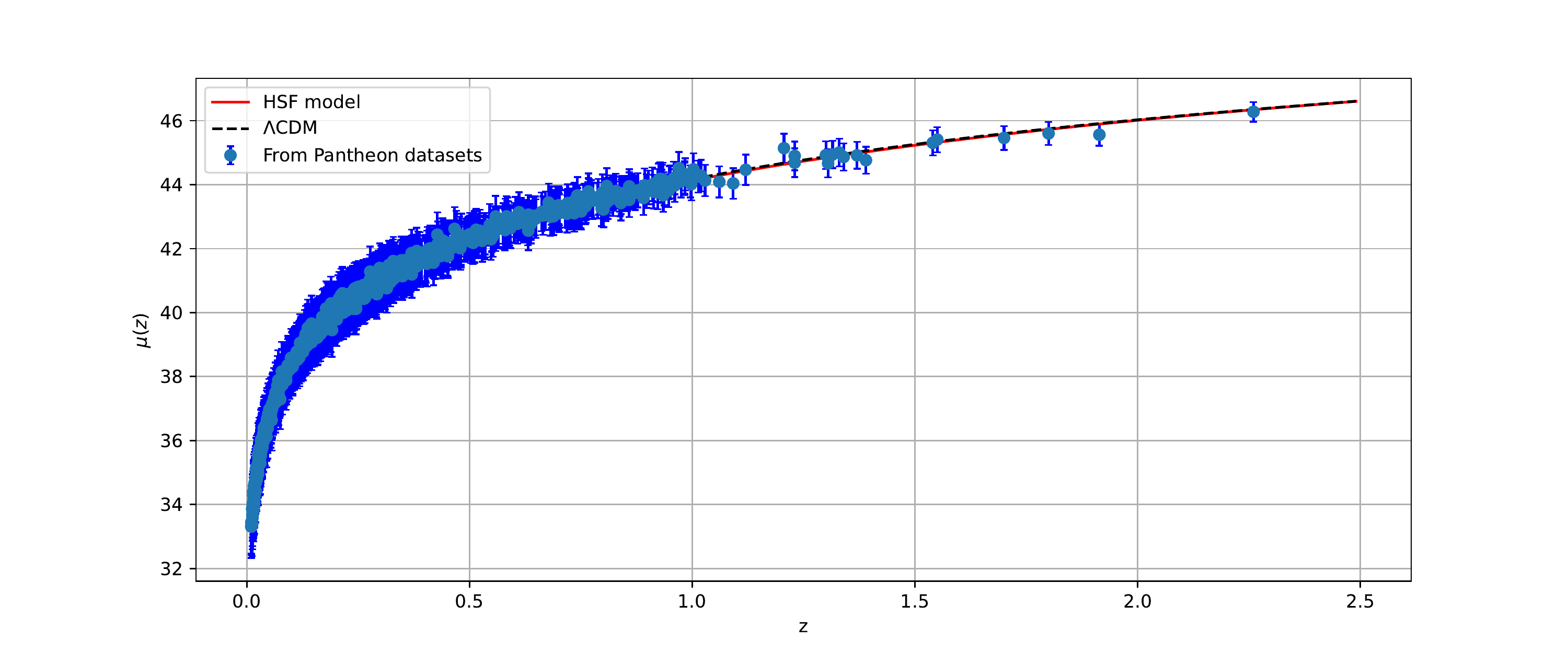}}
\caption{Evolution of distance modulus $\mu(z)$ with regard to red-shift $z$. The black dashed line represents the $\Lambda$CDM model, the red line is the HSF model curve, and the blue dots show error bars.}
\label{ErrorSNe}
\end{figure}

\begin{figure}[h]
\centerline{\includegraphics[scale=0.7]{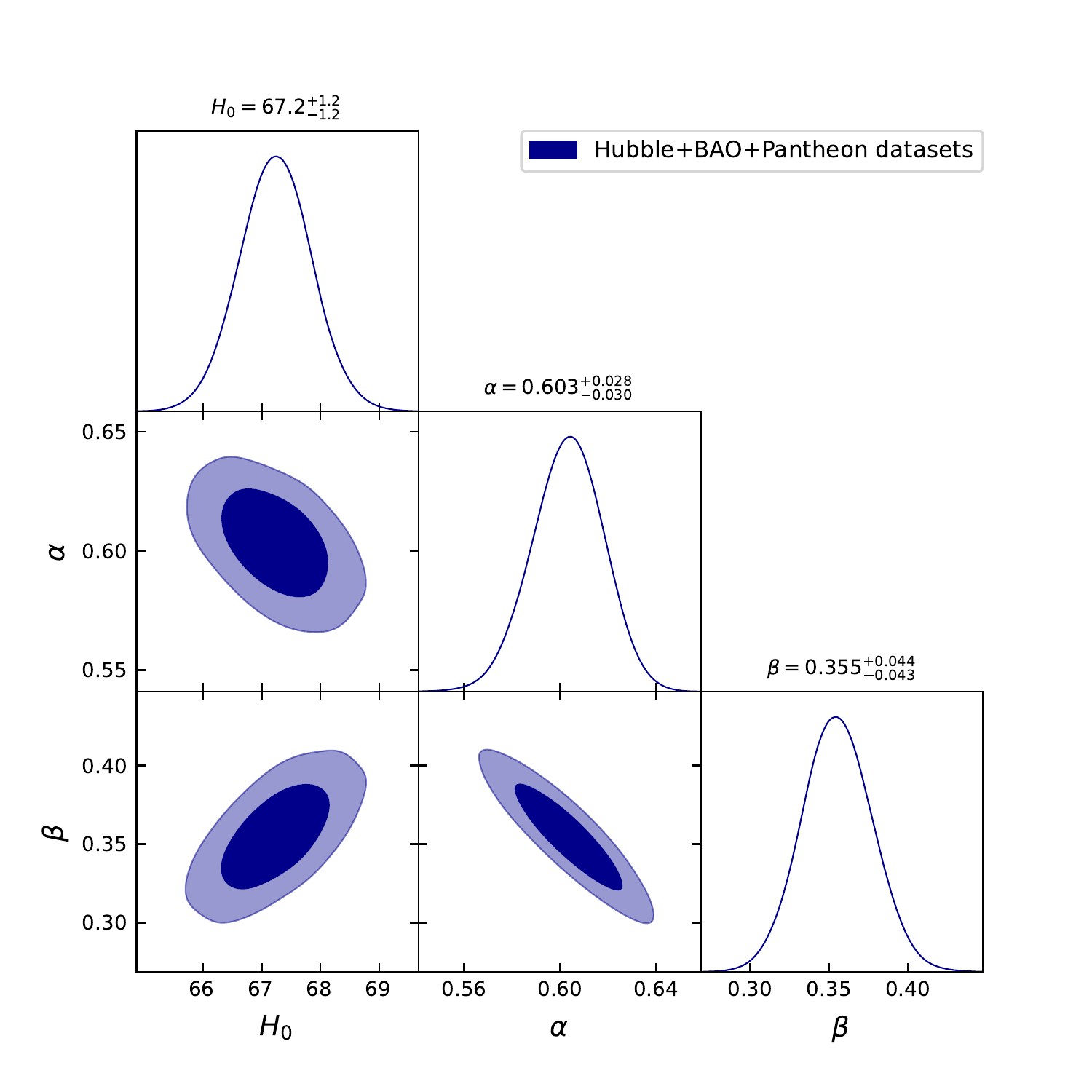}}
\caption{The $1-\sigma$ and $2-\sigma$ confidence curves for the model parameters $H_{0}$, $\alpha$, and $\beta$ with combined $Hubble+BAO+Pantheon$ data.}
\label{F_total}
\end{figure}

\begin{table*}[!htb]
\begin{center}
\renewcommand{\arraystretch}{1.5}
\begin{tabular}{l c c c c c c c c c}
\hline 
Data  & $H_{0}$ & $\alpha$ & $\beta$ & $q_{0}$ & $z_{tr}$ & $\omega_{0}$\\
\hline
$Priors$ & $(60,80)$  & $(0,10)$  & $(0,10)$ & $-$ & $-$ & $-$\\
$Hubble$   & $67.7^{+1.7}_{-1.7}$  & $0.57^{+0.14}_{-0.13}$  & $0.38^{+0.12}_{-0.13}$ & $-0.37^{+0.15}_{-0.15}$ & $0.83^{+1.87}_{-1.87}$ & $-0.61^{+0.14}_{-0.13}$\\
$Hubble+BAO+Pantheon$   & $67.2^{+1.2}_{-1.2}$  & $0.603^{+0.028}_{-0.030}$  & $0.355^{+0.044}_{-0.043}$ & $-0.343^{+0.062}_{-0.061}$ & $0.75^{+0.36}_{-0.37}$ & $-0.59^{+0.02}_{-0.03}$\\
\hline
\end{tabular}
\caption{Best-fit values of parameter space using $Hubble$ and $Hubble+BAO+Pantheon$ data.}
\label{tab}
\end{center}
\end{table*}
\end{widetext}

\section{The cosmological parameters}\label{sec5}
To understand the evolutionary behaviour of the Universe, we will examine the behavior of the EoS parameter as well as the behavior of energy conditions. In light of the enigmatic nature of DE and its uncertain nature, many candidates have been put forward. Among different approaches to understand the DE, the
cosmological constant $(\Lambda)$, has been considered to be the best and most straightforward. As a result, it is suggested that the acceleration of the Universe with the equation of state $\omega=-1$ is due to its repellent nature. Nevertheless, cosmological constants and constant EoS parameter afflict this genuine possibility.
The Universe is divided into two distinct phases: radiation-dominated phase $
\omega=\frac{1}{3}$ and matter-dominated phase $\omega=0$. It is well known
that the EoS parameter $\omega$ is essential for characterising the various energy-dominated processes in the evolution of Universe. Either the quintessence phase $-1 < \omega < 0$ or the phantom phase $\omega < -1$ can be used to anticipate the current state of the Universe. DE energy has also been investigated by determining the value of EoS parameter \cite{Carroll2003,Gong2007,Huang2008}.

\begin{figure}[h]
\centerline{\includegraphics[scale=0.72]{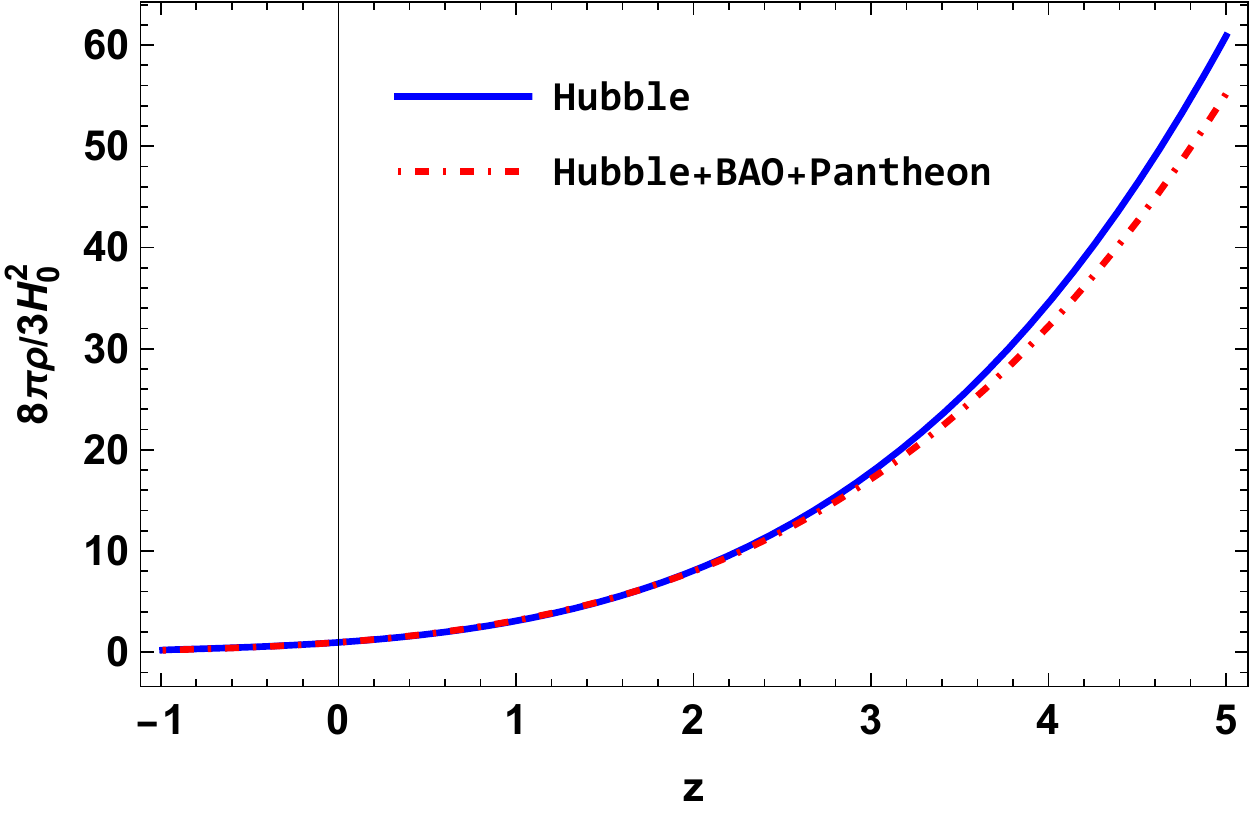}}
\caption{Evolution of $\rho$ versus $z$ with $b=0.2$ and $\gamma=1.5$.}
\label{F_r}
\end{figure}

\begin{figure}[h]
\centerline{\includegraphics[scale=0.72]{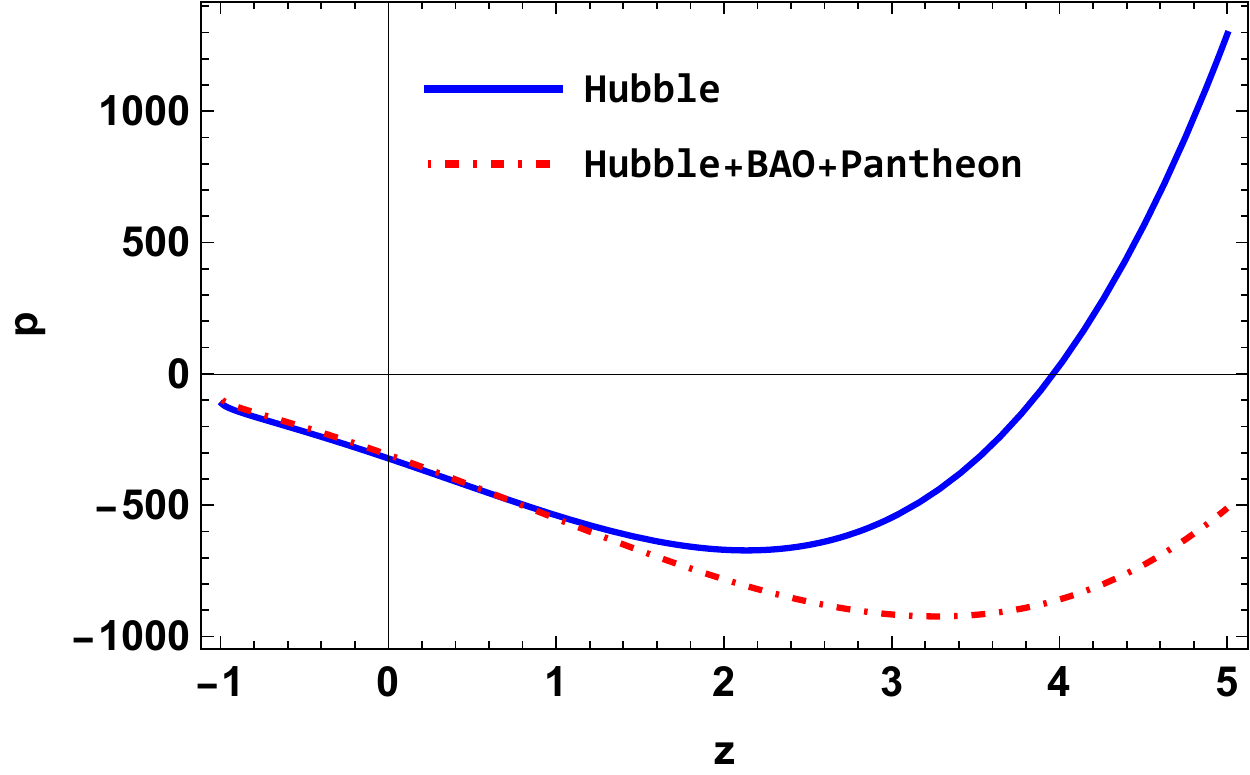}}
\caption{Evolution of $p$ versus $z$ with $b=0.2$ and $\gamma=1.5$.}
\label{F_p}
\end{figure}

\begin{figure}[h]
\centerline{\includegraphics[scale=0.72]{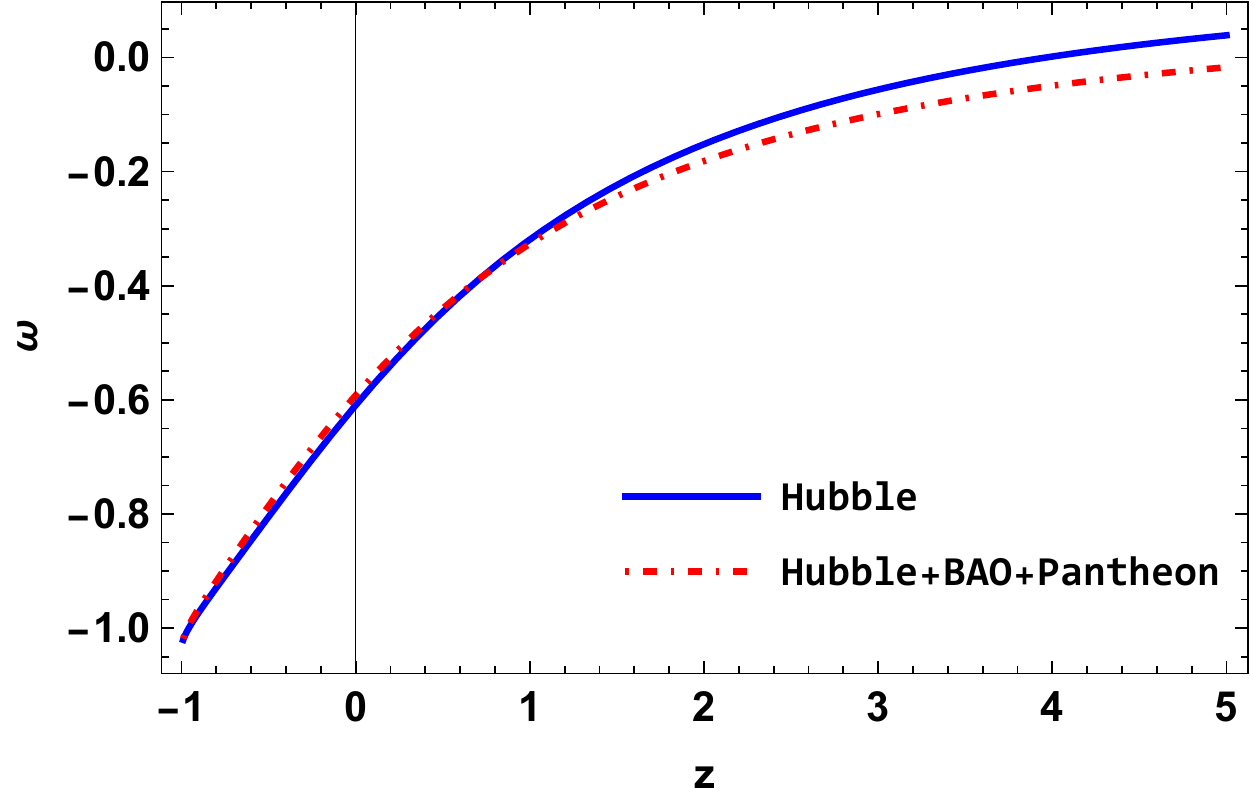}}
\caption{Evolution of $\omega$ versus $z$ with $b=0.2$ and $\gamma=1.5$.}
\label{F_w}
\end{figure}

The energy density shows the decreasing behaviour and remains positive throughout the evolution [Ref. FIG.- \ref{F_r}], whereas pressure gives negative behaviour at late time [Ref. FIG.- \ref{F_p}]. From FIG.- \ref{F_w}, one can see that the EoS parameter gives present value $\omega_{0}=-0.61^{+0.14}_{-0.13}$ and $\omega_{0}=-0.59^{+0.02}_{-0.03}$ \cite{Bonilla2018, Mukherjee2020, Brout2022} respectively for $Hubble$ and $Hubble+BAO+Pantheon$ dataset. For both the datasets, EoS shows quintessence phase at present and $\Lambda$CDM behaviour at late epoch.

Energy conditions are known to represent paths for implementing positive stress-energy tensors in the presence of matter. A fundamental casual and geodesic structure of space-time is characterized by the energy conditions, which can be used to characterize attractiveness of gravity. The cosmic evolution can also be influenced by the energy conditions, especially the acceleration and deceleration \cite{Capozziello2019}. Both classical and quantum instabilities are triggered when these energy conditions are violated \cite{Carroll2003}. We shall analyse the behaviour of weak, null, dominant and strong, energy conditions for the $f(Q,T)$ gravity model respectively abbreviated as WEC, NEC, DEC and SEC. The expressions for WEC, NEC, DEC, SEC are respectively $\rho +p\geq0,~\rho\ge0$, $\rho +p\geq0$, $\rho -p\geq0$, $\rho +3p\geq0$. The SEC must be violated for the actual acceleration phase of the our Universe. This limitation, together with the values of Hubble and deceleration parameter, one can test the viability of the model. In terms of Hubble parameter, we find the expressions of energy conditions as,
\begin{eqnarray}
\rho +p &=& -\frac{2 \left(\overset{.}{H}+3 (H-H_{y})^2\right)}{b+8 \pi}, \\
\rho -p &=& \frac{\overset{.}{H}+3 H^2}{b+4 \pi }, \\
\rho +3p &=& -\frac{\overset{.}{H}+3 H^2}{b+4 \pi }-\frac{4 \left(\overset{.}{H}+3 (H-H_{y})^2\right)}{b+8 \pi }.  \notag \\
\end{eqnarray}

\begin{figure}[]
\centerline{\includegraphics[scale=0.72]{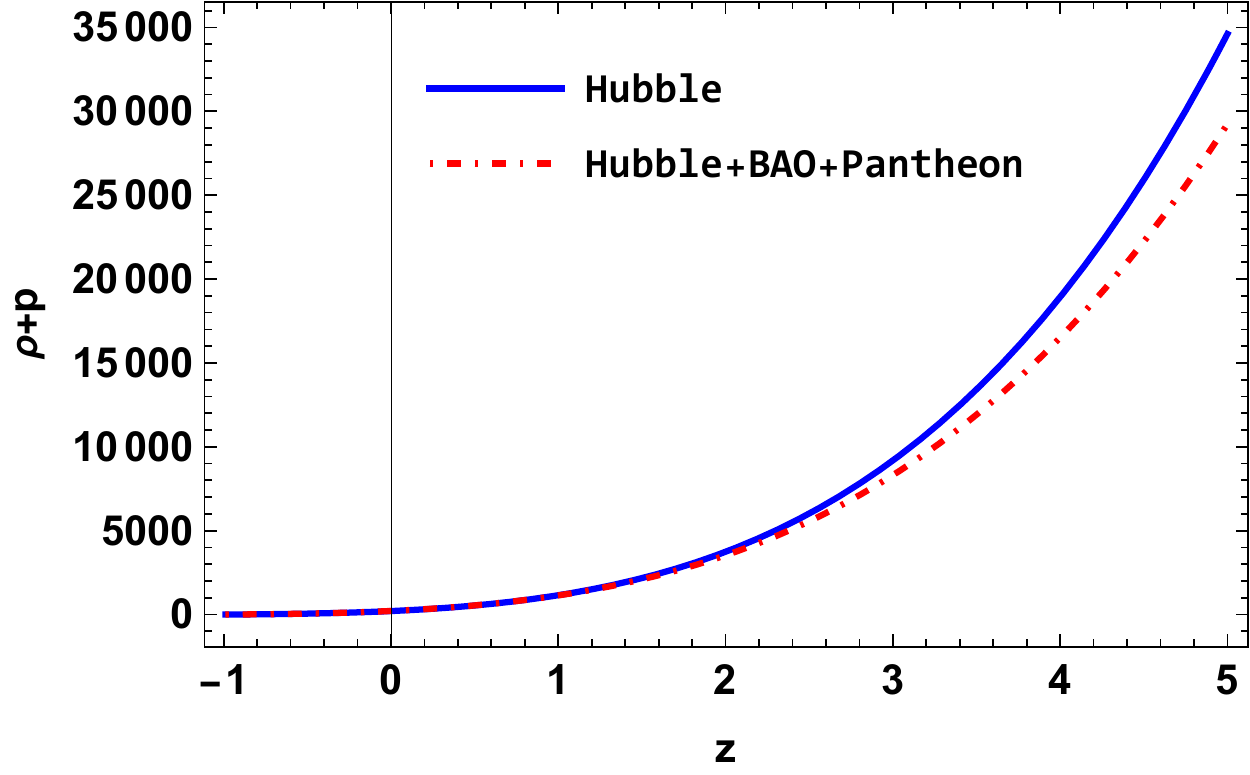}}
\caption{Evolution of $\rho+p$ versus $z$ with $b=0.2$ and $\gamma=1.5$.}
\label{F_NEC}
\end{figure}

\begin{figure}[]
\centerline{\includegraphics[scale=0.72]{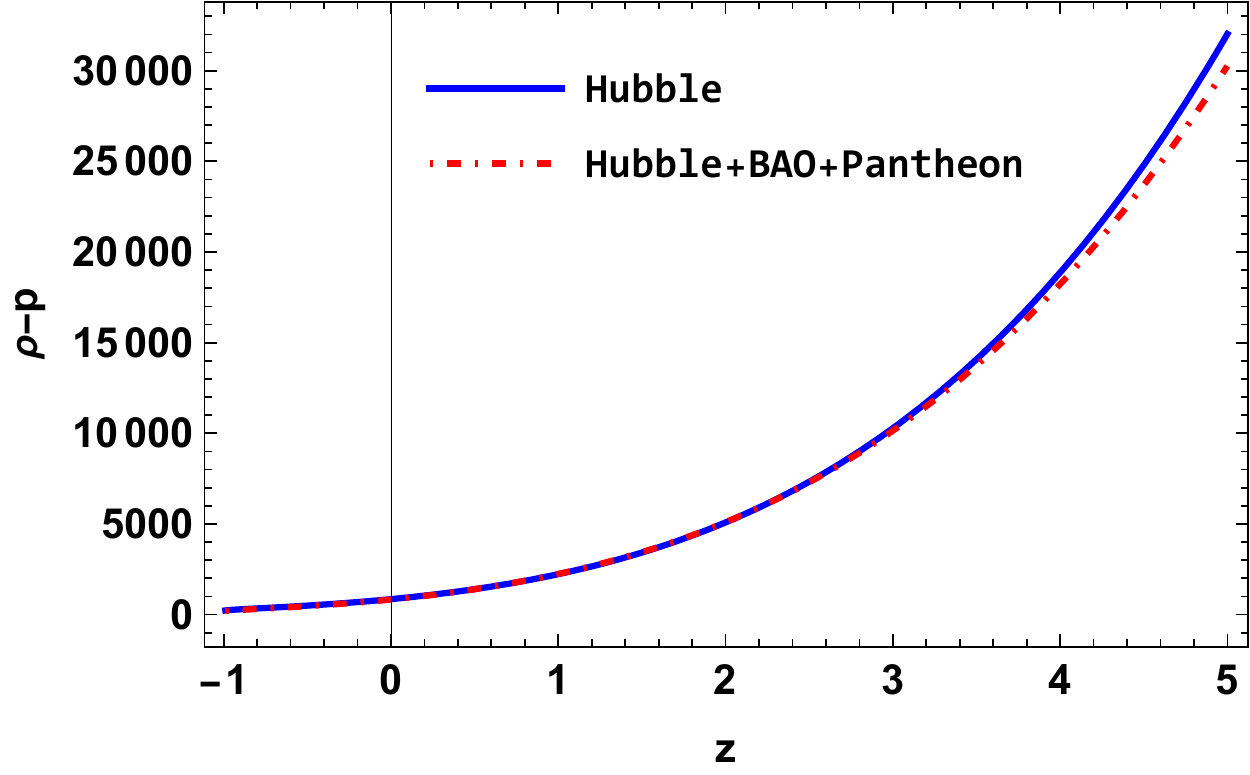}}
\caption{Evolution of $\rho-p$ versus $z$ with $b=0.2$ and $\gamma=1.5$.}
\label{F_DEC}
\end{figure}

\begin{figure}[]
\centerline{\includegraphics[scale=0.72]{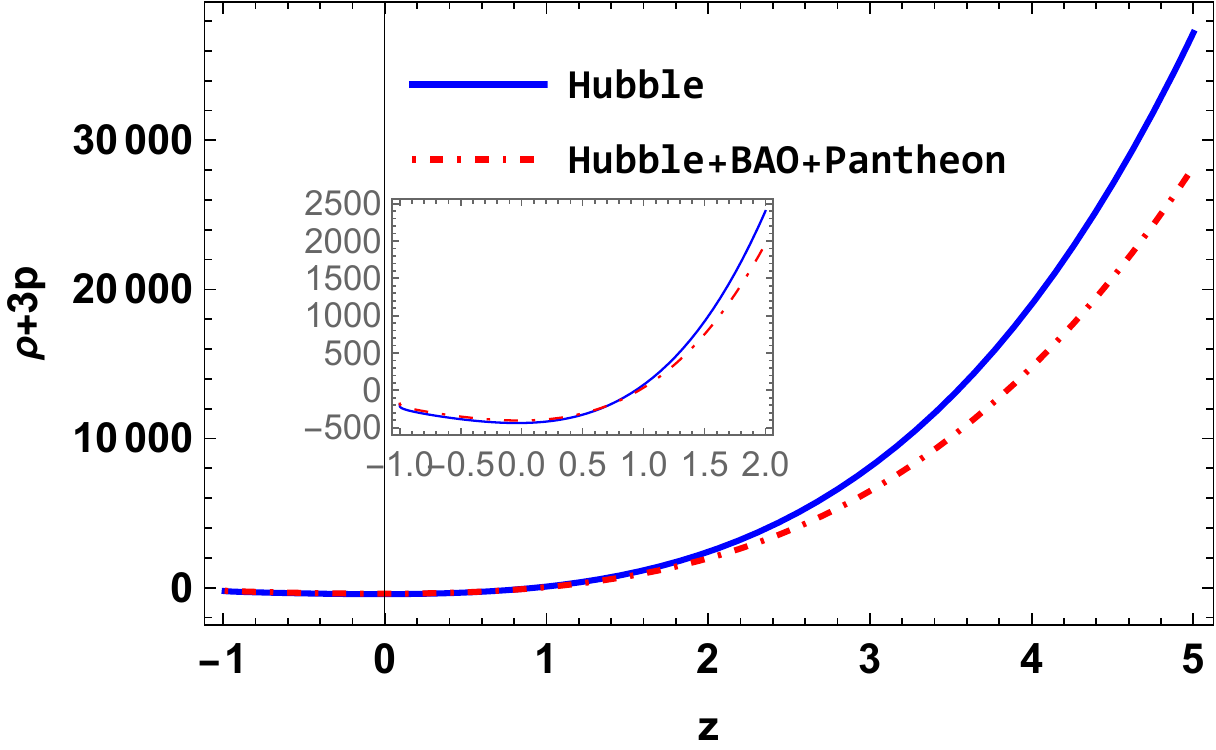}}
\caption{Evolution of $\rho+3p$ versus $z$ with $b=0.2$ and $\gamma=1.5$.}
\label{F_SEC}
\end{figure}

The behaviour of the NEC, DEC and SEC for $Hubble$ and $Hubble+BAO+Pantheon$ dataset has been presented in FIG.- \ref{F_NEC}, FIG.- \ref{F_DEC} and FIG.- \ref{F_SEC} respectively. The NEC decreases and then vanishes from early time to late time whereas the DEC, as expected, remains positive throughout the evolution. However, the SEC shows the transient behaviour, its evolution started from positive side at early time decrease gradually and crosses to negative side. Hence the violation of SEC has been observed at present and late times of the evolution. 

\section{Conclusion} \label{sec6} 
In the non-metricity based gravitational theory, the $f(Q,T)$ gravity, we have presented an accelerating cosmological model of the Universe in a flat and anisotropic background. A well motivated form of the function $f(Q,T)$ has been used such that under certain condition, it can reduce to GR. The parameters in the model have been constrained through different cosmological data sets. Mostly, we have focused on $Hubble$, $BAO$ and $Pantheon$ observational datasets. The free parameters of the Hubble parameter obtained from HSF has been constrained in a range of confidence between $1-\sigma$ and $2-\sigma$, the details has been provided in TABLE-\ref{tab}. The error bar for $Hubble$ data set shows that the considered $H(z)$ have same background as $\Lambda$CDM at small redshift and diverges from $\Lambda$CDM at high
redshift [FIG.- \ref{ErrorHubble}]. Whereas the error bar plot for $pantheon$
data set shows that the $H(z)$ passes through middle of the light curves
[FIG.- \ref{ErrorSNe}].

As mentioned earlier, the HSF is the combination of power law and exponential functions and can describe the early inflationary phase and late time acceleration phase. The transition from deceleration to acceleration for $Hubble$ and
$Hubble+BAO+Pantheon$ respectively shows the value $z_{t}=0.83^{+1.87}_{-1.87}$ and $z_{t}=0.75^{+0.36}_{-0.37}$. The present value of the deceleration parameter has been obtained as, $q_{0}=-0.37^{+0.15}_{-0.15}$, and $q_{0}=-0.343^{+0.062}_{-0.061}$, respectively for the $Hubble$ and $Hubble+BAO+Pantheon$ datasets. The behaviour of deceleration parameter $q<0$ in late time shows the acceleration of the Universe. Further the dynamical parameters are assessed. The energy density remains positive throughout the evolution and shows a decreasing behavior. The present values of the EoS parameter are obtained as, $
\omega_{0}=-0.61^{+0.14}_{-0.13}$ and $\omega_{0}=-0.59^{+0.02}_{-0.03}$ for $Hubble$ and $Hubble+BAO+Pantheon$ datasets respectively. Accordingly at present the quintessence behavior has been noted and at late times it approaches $\Lambda$CDM behavior.

The verification of the behaviour of energy conditions of the model has been inevitable in the context of modified gravity. The violation of SEC at late times further validates the model. As usual the the DEC does not violate and the NEC also remains positive throughout, though vanishes at the end. The SEC is violated, which is consistent with the accelerated expansion of the universe. This violation may indicate the existence of fundamental physics in the late time universe that requires modifications to the current understanding of gravity. Though we have not performed the stability analysis of the model in $f(Q,T)$ gravity, but the cosmological behaviours pertaining to different parameters shows its consistency with the geometry. This study may provide additional theoretical considerations for the better understand on the late time behaviour of the Universe in the non-metricity based gravitational theories.

\section*{Acknowledgments:}BM acknowledges IUCAA, Pune, India, for hospitality and support during an academic visit where a part of this work has been accomplished. The authors are thankful to the anonymous reviewer for the comments and suggestions to improve the quality of the paper.

\textbf{Data availability} There are no new data associated with this
article.

\end{document}